# Watch an AI Weather Model Learn (and Overshoot) Tropical Cyclone Intensity

**Rebecca Baiman[1,2], Ankur Mahesh[3,4], and Elizabeth A. Barnes[1,2,5]**

[1]Department of Atmospheric Science, Colorado State University, Fort Collins, CO, USA.

[2]Faculty of Computing and Data Science, Boston University, Boston, MA, USA

[3]Earth and Environmental Sciences Area, Lawrence Berkeley National Laboratory (LBNL), Berkeley, CA, USA.

[4]Department of Earth and Planetary Science, University of California, Berkeley, CA, USA

[5]Department of Earth and Environment, Boston University, Boston, MA, USA

Corresponding author: Rebecca Baiman (rbaiman.earth@gmail.com)

**Key Points:**

- We introduce the use of task-specific training dynamics for “watching” an AI weather emulator learn extreme weather events.
- For a subset of tropical cyclones, a Spherical Fourier Neural Operator overpredicts storm intensity resulting in a loss of skill.
- Reducing environmental moisture eliminates this degradation in skill, confirming moisture as a driver of these intensity errors.

## Abstract

In a changing climate, artificial intelligence (AI) weather models have the potential to provide cheaper, faster, and more accurate forecasts of high-impact weather events. To realize this potential and gauge trustworthiness, there is a need for more research on how models learn extreme events and how that learning might be improved. Here, we investigate how a Spherical Fourier Neural Operator (SFNO) learns tropical cyclones (TCs) by saving every checkpoint from training and analyzing storm-specific metrics. We find that a subset of storms show early TC intensity forecast skill that degrades into systematic overprediction in the second half of training. Moisture perturbation experiments show a causal relationship between anomalously moist environments and this skill degradation. This work provides a first example of leveraging task-specific training dynamics to further our understanding of how AI weather models learn extreme events.

## Plain Language Summary

A new frontier of weather forecasting uses artificial intelligence (AI) instead of traditional physics-based models to make forecasts. These AI models have the potential to be cheaper, faster, and more accurate than traditional models. However, AI weather models still have room for improvement in forecasting extreme weather events like tropical cyclones. We look at how one of these machine learning models learns tropical cyclones during training of the model. We find that for some tropical cyclones, the model can produce a better forecast of storm intensity early in its training that gets worse in the second half of training.

## 1 Introduction

As a warming climate drives more frequent and more extreme weather events (Intergovernmental Panel On Climate Change (IPCC), 2023), artificial intelligence (AI) emulators open the door to faster, cheaper, and potentially more accurate weather forecasts compared to physics-based models (Pasche et al., 2025). These emulators, including but not limited to Google's GraphCast (Lam et al., 2023), Huawei's PanguWeather (Bi et al., 2023), and NVIDIA's FourCastNetV2 (Bonev et al., 2023), are trained on European Center for Medium-Range Weather Forecasts Reanalysis v5 (ERA5) and use variations of mean squared error or mean absolute error as their loss function. Since skillful forecasting of extreme weather is integral to the utility of these emulators, evaluation based on average forecast error is inadequate (Olivetti & Messori, 2024a).

A growing body of research addresses the skill of these AI emulators in forecasting extreme weather (Charlton-Perez et al., 2024; Olivetti & Messori, 2024a, 2024b; Pasche et al., 2025; Sun et al., 2025). For example, DeMaria et al. (2025) found that AI emulators capture seven-day Tropical Cyclone (TC) tracks with comparable skill to operational forecast models, but consistently and substantially underestimate the cyclone intensity. The use of ensembles, particularly huge ensembles, could improve the ability of these emulators to capture extreme events (Eyring et al., 2024; Mahesh et al., 2025c). Along with increased ensemble size, some argue for focusing the training or post-processing of AI emulators on extremes (Olivetti & Messori, 2024a).

Missing from this conversation, and crucial to gauging trust in the use of AI for forecasting extreme weather, is an understanding of *how* AI emulators currently learn extreme weather. Sun et al. (2025) begins to explore this topic using retraining experiments. They find that emulators struggle to extrapolate to stronger TCs from weaker storms in training data, but there is evidence of learning via translocation, i.e., learning from similarly strong storms in other regions. Their focus is the model's ability to extrapolate to out-of-distribution extreme events, but their result on learning via translocation speaks more generally to how models learn extreme weather.

Here, we examine how models learn in-distribution extreme weather through training. We introduce a computationally light diagnostic approach: examining how certain metrics of extreme weather, in this case TCs, evolve through training. This work follows from the field of training dynamics in deep learning, which is garnering particular attention for its application to large language models (Betley et al., 2026; Kangaslahti et al., 2026; Tirumala et al., 2022). The most basic example of training dynamics is plotting the training and validation loss curve, but the general loss curve obscures changes in model behavior with respect to certain tasks (Hu et al., 2025). While the model's weights are updated according to a globally averaged loss function, one can define a particular task (e.g. location of a TC) and evaluate the training dynamics of this metric. Additionally, training dynamics can be evaluated for a subset of samples and clustered according to their learning trajectory to identify common behaviors (Kangaslahti et al., 2026). Here, we take a first look at the training dynamics of extreme weather in an AI emulator. Specifically, we evaluate the learning trajectories of 106 TCs in a spherical Fourier neural operator (SFNO).

## 2 Methods

The model and subsequent analysis uses ERA5 (Hersbach et al., 2020) at a 0.25 degree latitude-longitude resolution. In accordance with the AGU publication guidelines, we note the use of VS Code Copilot for assistance in data analysis and sentence structure suggestions. No code from Copilot was implemented without individual line review and no text was copied directly from Copilot.

### 2.1 Model

We use a Spherical Fourier Neural Operator (SFNO) model, the same architecture used in AI2 Climate Emulator (ACE2) (Watt-Meyer et al., 2025) and NVIDIA's FourCastNet V2 (Bonev et al., 2023). We use the framework from Mahesh et al. (2025b) huge ensemble, with a scale factor of 3 and an embedding size of 384. The scale factor determines how much the input is downsampled horizontally in the SFNO, while the embedding size sets how large the internal feature representation is within the model's state space. The model is built from the open source SFNO v0.1.0 in the modulus-makani Python repository (NVIDIA, n.d.). The SFNO model includes an encoder (a dense neural network that maps each latitude-longitude grid cell to a higher dimensional latent space), 8 SFNO blocks (spherical convolutions applied in the spectral domain followed by a dense neural network), and a decoder (a dense neural network that maps the latent space back into the physical space). SFNOs respect spherical geometry and retain physically plausible dynamics (Bonev et al., 2023). See Bonev et al. (2023) Figure 3 for a schematic of this architecture and Mahesh et al. (2025b) for a complete model description and discussion of hyperparameters, such as scale factor and embedding dimension.

Training data includes 6-hourly ERA5 reanalysis from 1979–2015 (inclusive) for 74 variables and 13 pressure levels for relevant variables (Mahesh et al., 2025b). The cosine of the solar zenith angle, orography, and land-sea mask are included as non-prognostic inputs. The training of this SFNO model is identical to the training of a single ensemble member in Mahesh et al. (2025b) with two additions: 1) After 70 epochs of training with a cosine annealing learning rate scheduler starting at 1e-3, we include a 20 epoch 2-step fine-tuning with a cosine annealing learning rate scheduler starting at 1e-4. 2) Model weights are saved after each epoch of training. We refer to these saved states as checkpoints, where checkpoint n corresponds to the model weights saved after epoch n.

The model is initialized with random weights and trained to minimize a mean-squared error between forecasts and ERA5 training data. Each epoch of training iterates through all training timesteps and the model is trained to forecast the next 6 hour timestep. All other details of training and hyperparameter selection are found in Mahesh et al. (2025b). The model's overall best checkpoint evaluated on validation years 2016 and 2017 is checkpoint 70 without fine-tuning, and checkpoint 89 with fine-tuning. We reference both of these checkpoints when analyzing TC forecast errors.

### 2.2 Tropical Cyclone Forecasts

We select all TCs (category 1–5) in the Northwest Pacific and North Atlantic basins that occurred in five years of available data outside the training set (2016, 2017, 2019, 2021, and 2022) (Knapp et al., 2010). TCs in validation years 2016 and 2017 are included, which could slightly inflate skill for these storms. However, the model's loss function is based on global MSE so we expect this leakage is negligible and well-worth the increased number of TC cases for analysis. The collection of 106 TCs (39 in the Atlantic and 67 in the Pacific) includes 20 Category 5, 20 Category 4, 15 Category 3, 20 Category 2, and 31 Category 1 storms. We run five-day forecasts for each storm where, crucially, the valid timestep is the observed peak intensity of the storm defined by the International Best Track Archive for Climate Stewardship (IBTrACS) (Knapp et al., 2010).

We choose a five-day lead time to focus on synoptic-scale features that drive TCs. We run a five-day (20 timestep rollout) SFNO inference for each TC and repeat this inference run with model checkpoints from each of the 90 training epochs. This results in 90 forecasts for each of the 106 TCs. We compare forecasted values to "true" ERA5 values.

### 2.3 Assessment of Training Dynamics

We define a storm-relative domain consisting of a 1200 km radius circle around the minimum MSLP value in the ERA5 valid time (e.g. Figure 1a). This domain is unique for each storm but constant through checkpoints. 1200 km is chosen to represent approximately $\pm 1$ day of movement of the TC.

Within this domain, we assess the valid time TC forecasts using two metrics: intensity and location. The intensity error of the TC is evaluated using the difference between the forecasted minimum MSLP value and the ERA5 minimum MSLP. The location error of the TC is evaluated as the distance between the forecasted minimum MSLP and the true minimum MSLP. Note that the maximum location error is 1200 km as set by the radius of the storm-relative domain so this metric has a discrete upper bound unlike MSLP.

For each storm, we calculate storm intensity and location metrics at each of the 90 checkpoints. We define "learning trajectory" as the pattern of a metric over the 90 checkpoints. By analyzing all 106 TC learning trajectories, we can evaluate the TC-specific training dynamics of the SFNO. In other words, these metrics allow us to "watch" the model learn TCs through training.

First, we evaluate learning trajectories of Hurricane Larry over 90 checkpoints using our two error metrics (Figure 1). Hurricane Larry is used as an illustrative case study to introduce storm-specific learning trajectories and demonstrate that intensity and location errors can diverge through training.

When we expand this to all 106 TC cases, we implement a k-means clustering to group storms into three TC intensity learning trajectory patterns (Figure 2a). After standardizing the data, we use the elbow method and silhouette scores and find that 2 or 3 clusters are reasonable for this dataset (Figure S1a and b). We choose 3 clusters despite a slightly lower silhouette score to identify an additional learning trajectory group (see Figure S1, Text S1 for details on k-means clustering). Next, we evaluate how skillful each cluster is in forecasting TC intensity with the absolute error in minimum MSLP (Figure 2b). To identify trends in TC intensity error over training, we apply a Gaussian (sigma = 2) filter along each TC learning trajectory (Figure 2b). Note that the minimum in smoothed learning trajectories does not necessarily represent the singular best performing checkpoint. Instead, it represents the point of training at which the model sustained the lowest error over multiple epochs (e.g. Figure 2c).

### 2.4 Moisture Experiments

To identify a causal relationship between anomalously moist environments and learning trajectory trends, we conduct reduced moisture experiments. First, we try reducing moisture just in the initial input data but the model's autoregressive rollout restores moisture to near-original levels within the first few rollout steps, effectively negating the perturbation. Instead, we reduce global total column water vapor and all levels of specific humidity in the initial input data *and* for each 6 hour rollout. We test a range of moisture reductions and select the values that reduce forecast moisture to near the all-storm average throughout the five-day forecast (Figure S2a). Experiment 1 reduces global moisture by 2% in the initial input data and 1% in the rollout, which results in a decrease in moisture but maintains a storm environment more moist than average storm moisture. Experiment 2 reduces moisture by 3% in the initial input data and 2% in the rollout, resulting in average domain moisture very close to the all-storm average (Figure S2a).

## 3 Results

### 3.1 Hurricane Larry Learning Trajectory

Hurricane Larry began as a tropical depression in the eastern Atlantic on August 31st and traveled northwest reaching its maximum intensity on September 4th, 2021 at 6 UTC as a category 3 storm. Consistent with the methodology applied to all TCs, the Hurricane Larry forecasts were initialized five days prior to its peak intensity. This initial timestep is 12 hours before the system was classified as a tropical depression.

Figure 1 shows the location and intensity errors for the five-day Hurricane Larry forecasts through 90 checkpoints of training. Note that the variance in both metrics decreases to near zero by checkpoint 70 and then again towards checkpoint 90, which is consistent with the cosine annealing learning rate applied to the first 70 epochs and the 20 checkpoints of fine-tuning. In the first two checkpoints of training the SFNO does not forecast a TC within 1200 km of the true location, but by checkpoint 5 the forecasted TC is within 400 km of the truth (Figure 1a). Hurricane Larry location error decreases through training with checkpoints 65–70, 89, and 90 lying within 100 km of the true location (dark pink line, Figure 1b).

Unlike location error, Hurricane Larry intensity error increases through the second half of training. At checkpoint 30, the intensity of the hurricane is skillfully forecasted with an error of 0.76 hPa, but the model’s best overall checkpoints 70 and 89 have intensity errors larger than 18 hPa (Figure 1a). This difference is the result of an increasing trend in Hurricane Larry intensity error over the second half of training (black line in Figure 1). The forecast minus truth for Checkpoints 30, 50, 70, and 90 is less than zero around the forecasted storm center (blue shading in Figure 1a) while minimum MSLP absolute error increases, demonstrating that the forecasted minimum MSLP is  too low. Thus, the degradation in forecast skill for Hurricane Larry’s intensity in the second half of training is attributable to the SFNO *overpredicting* the intensity of this storm.

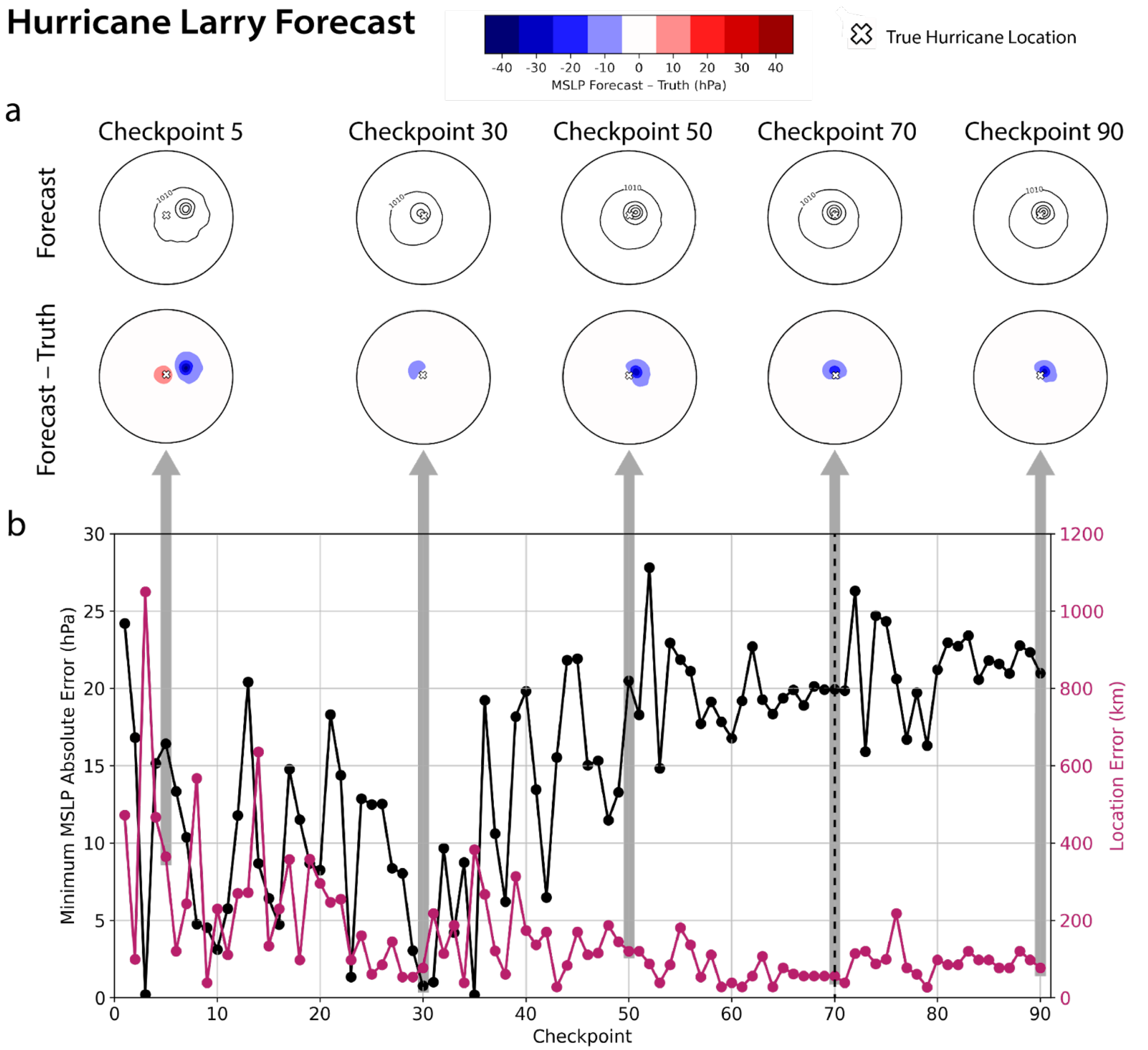

**Figure 1** Training dynamics of Hurricane Larry. Five-day forecast of Hurricane Larry for checkpoints 5, 30, 50, 70, and 90 centered on the valid time (September 4th, 2021 at 6UTC) storm center (white x) (**a**). Black contours show forecast MSLP every 10 hPa, starting at 1010 hPa. Shading shows SFNO forecast minus ERA5 MSLP, where blue indicates forecasted MSLP is lower than truth. Minimum MSLP absolute error in hPa (black) and location error in km (dark pink) for all 90 training checkpoints (**b**). The dashed vertical line at checkpoint 70 marks the start of 2-step fine-tuning.

**3.2 TC Intensity Training Dynamics**

Moving from the case study of Hurricane Larry to a consideration of 106 TCs, we first ask if and when in training the SFNO can successfully forecast TCs. Only 42% of the storm cases feature a closed MSLP low, defined as a pressure minimum surrounded by at least one closed isobar in the domain, in the ERA5 initialization data within the storm domain. In the remaining cases, the SFNO must successfully forecast the movement of a storm into the domain or the onset of a storm. Impressively, by checkpoint 5, 75% of the five-day forecasts produce a closed MSLP low (not shown). The model's best checkpoint 70 forecasts a closed MSLP low in the storm domain for 92% of cases. Because our intensity error metric is a function of forecasted minimum MSLP, this analysis confirms that for the majority of storms minimum MSLP represents the intensity of a storm center rather than the edge of a regional MSLP gradient.

After verifying our storm intensity metric, we plot the minimum MSLP forecast minus truth for all TCs over 90 checkpoints (thin lines, Figure 2a,b). We observe that for many storms, this learning trajectory decreases over training. Similar to Hurricane Larry, a portion of cases fall below zero through the second half of training indicating that the SFNO is overpredicting the intensity of these TCs. To identify patterns of learning, we apply k-means clustering to the minimum MSLP forecast minus truth learning trajectories from all 106 storms. The metrics used to determine the optimal number of clusters are detailed in Text S1 and Figure S1. This method partitions the storms into three clusters based on how their intensity error evolves through training (Figure 2a).

The SFNO consistently underpredicts storm intensity for Cluster 1 TCs: out of 2,700 individual forecasts (30 storms x 90 checkpoints), 98% have a positive forecast minus truth value. While most storm intensities are underpredicted for this cluster, there is a slight improvement in intensity forecasts through training. Checkpoints 1–10 have an average error of 18 hPa while checkpoints 60–70 and 80–90 have average errors of 13 hPa. At the best overall model (fine-tuning) checkpoint 70 (89), only 23% (33%) of cases have errors within 10 hPa of the truth.

Cluster 2 is the largest cluster, representing 51% of TC cases. Cluster 2 has an average learning trajectory similar to a typical loss curve: Forecast minus truth values start around 10 hPa and trend downward over training. Checkpoints 50–70 hover around zero and at checkpoint 70 (89), 87% (89%) of cases have errors within 10 hPa of the truth.

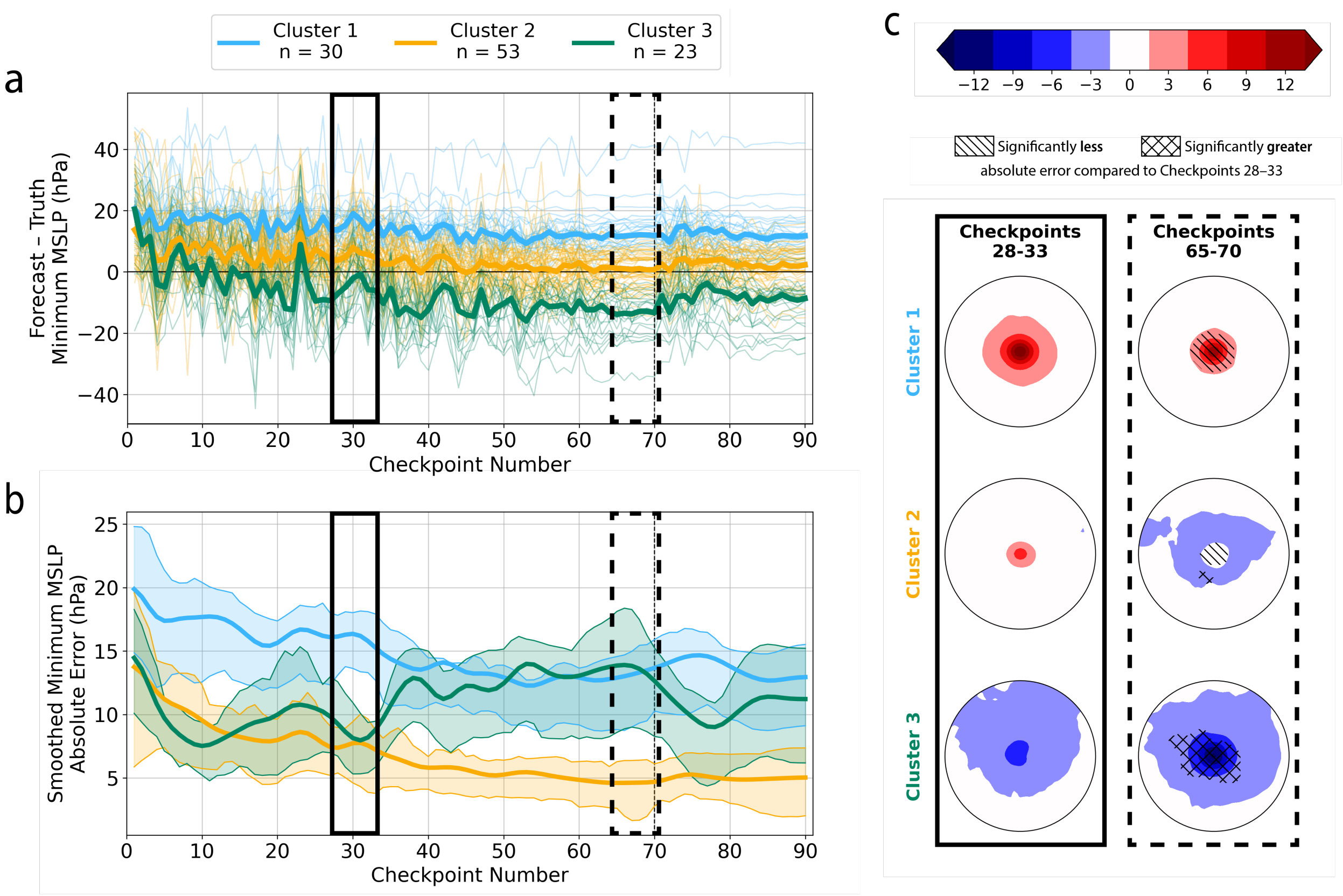

**Figure 2** Training dynamics of 106 TCs**.** TC forecast minus truth minimum MSLP for each storm (thin lines, **a**) and mean for each cluster (thick lines, **a**). Mean and interquartile range for each cluster of absolute value of forecast minus truth minimum MSLP smoothed along checkpoints (Gaussian filter, sigma = 2) (**b**). Black solid and dashed rectangles outline checkpoints 28–33 and 65–70 in a and b. Storm-centered minimum MSLP composites in clusters 1–3 (rows) at checkpoints 28–33 (left) and 65–70 (right) (shading, **c**). Significantly greater (smaller) values at checkpoint 67–70 determined by t test ($p<0.05$) where storm-centered MSLP forecast minus truth difference is at least 3 hPa (stripes and hatching, **c**).

Twenty-three TCs fall into Cluster 3, including Hurricane Larry. On average, the SFNO overestimates the intensity of Cluster 3 storms from checkpoint 30 onward (Figure 2a,b; bold green lines). This is true of individual storms in this cluster as well: 93% of the individual TC forecasts after checkpoint 29 are too intense. Similar to Cluster 1, only 30% of Cluster 3 checkpoint 70 forecasts fall within 10 hPa of the truth. Cluster 3 does better at checkpoint 89 with 56% of cases falling within 10 hPa of the truth.

To verify that the cluster learning trajectories represent meaningful differences in storm intensity beyond minimum MSLP, we plot the maximum 10m wind speed within each storm domain by cluster (Figure S3). Maximum 10m wind speeds are consistent with minimum MSLP values: Cluster 1 underestimates wind, Cluster 2 also underestimates wind but lower error on average, and Cluster 3 overestimates the strength of the wind in the second half of training.

To identify the learning trajectory trends, we take the minimum MSLP absolute error (absolute value of minimum MSLP forecast - truth) and smooth each TC case along the training checkpoints using a Gaussian filter with a sigma of 2. The mean and interquartile spread of these smoothed learning trajectories (Figure 2b) show distinct training dynamics for each cluster. Cluster 1 and 2 errors decrease over training while Cluster 3 has a minimum around checkpoint 30 with relatively low variability before a striking positive trend between checkpoint 30 and 70. Another increase in error is evident in the fine-tuning checkpoints 78–83. To highlight the unique behavior of the TC intensity learning trajectories for each cluster, we plot the storm-centered composites of MSLP absolute error for checkpoints 28–33 and checkpoints 65–70 (Figure 2c). We find that Cluster 1 and 2 have significantly lower MSLP errors around the storm center for checkpoints 65–70 compared to checkpoints 28–33. On the other hand, Cluster 3 has significantly higher MSLP error in checkpoints 65–70 compared to checkpoints 28–33 (Figure 2c). Cluster 3 storm intensity shows the same degradation in skill as the Hurricane Larry case study. From Figure 2a, we know that the Cluster 3 skill degradation towards the end of training is due to an overprediction of storm intensity.

### 3.3 TC Characteristics Associated with SFNO Overshooting Storm Intensity

The distinct behavior of Cluster 3 TC intensity through training motivates an investigation of what distinguishes this group of 23 storms. Is there some physical characteristic driving the SFNO's overprediction of Cluster 3 storm intensity and the resulting degradation in skill over the second half of training?

In initial investigations, we find that Cluster 3 is not distinguished by ERA5 TC characteristics including category or strength of the storm, storm track, location of the storm, storm intensification over forecast period, El Niño Southern Oscillation phase, year, day of year, or time of genesis relative to initial forecast date (Figure S4). However, there *is* a significant difference in the environmental moisture associated with Cluster 3 storms, as represented by 700 hPa specific humidity (Figure 3a). This is true in the ERA5 data at the valid timestep (Figure 3a) and throughout the five-day forecast period (Figure S5a). We note that on the day of initialization, Cluster 3 storms feature anomalous moisture co-located with the developing storm locations (Figure S5a, row 1). This is likely associated with warmer sea surface temperatures southeast of the storm (Figure S5b).

Because Cluster 3 storms are distinguished by anomalously moist environments, we ask whether the model's moisture forecasts drive the intensity overprediction. Plotting 700 hPa specific humidity in checkpoints 32 and 69 we find that the SFNO correctly forecasts enhanced moisture in Cluster 3 storms compared to the all-storm average (Figure 3a) in both checkpoints. Thus, the SFNO is successfully capturing the signal of enhanced moisture in Cluster 3 storms. We also see that the spatial pattern and magnitude of enhanced moisture in Cluster 3 is not meaningfully different in checkpoint 32 versus 69 (Figure 3a). From this analysis, along with the comparison of all checkpoints 28–33 and 65–70 (Figure S6), we conclude that there is *not* a meaningful difference in the forecast of specific humidity between checkpoints where the model is more successfully forecasting storm intensity (28–33) and where it overpredicts Cluster 3 storm intensity (65–70). Thus, any contribution of elevated moisture to Cluster 3 intensity overprediction is not due to the model misrepresenting moisture. The question remains if the elevated moisture plays a causal role in the overprediction of TC intensity and resulting skill degradation.

To address this question, we conduct two experiments in which we run five-day forecasts of Cluster 3 storms with reduced moisture environments. The first experiment results in reduced anomalous moisture (Figure 3b), while the second experiment eliminates anomalous moisture and introduces some anomalously dry areas (Figure 3c). The first experiment improves TC intensity forecasts, and the second experiment eliminates Cluster 3 skill degradation trend altogether (Figure 3d, Figure S2b). Reducing moisture improves TC intensity forecasts while maintaining similar skill in TC location error (Figure S2c). These experiments demonstrate a causal relationship between the elevated environmental moisture and the TC intensity skill degradation in Cluster 3 storms.

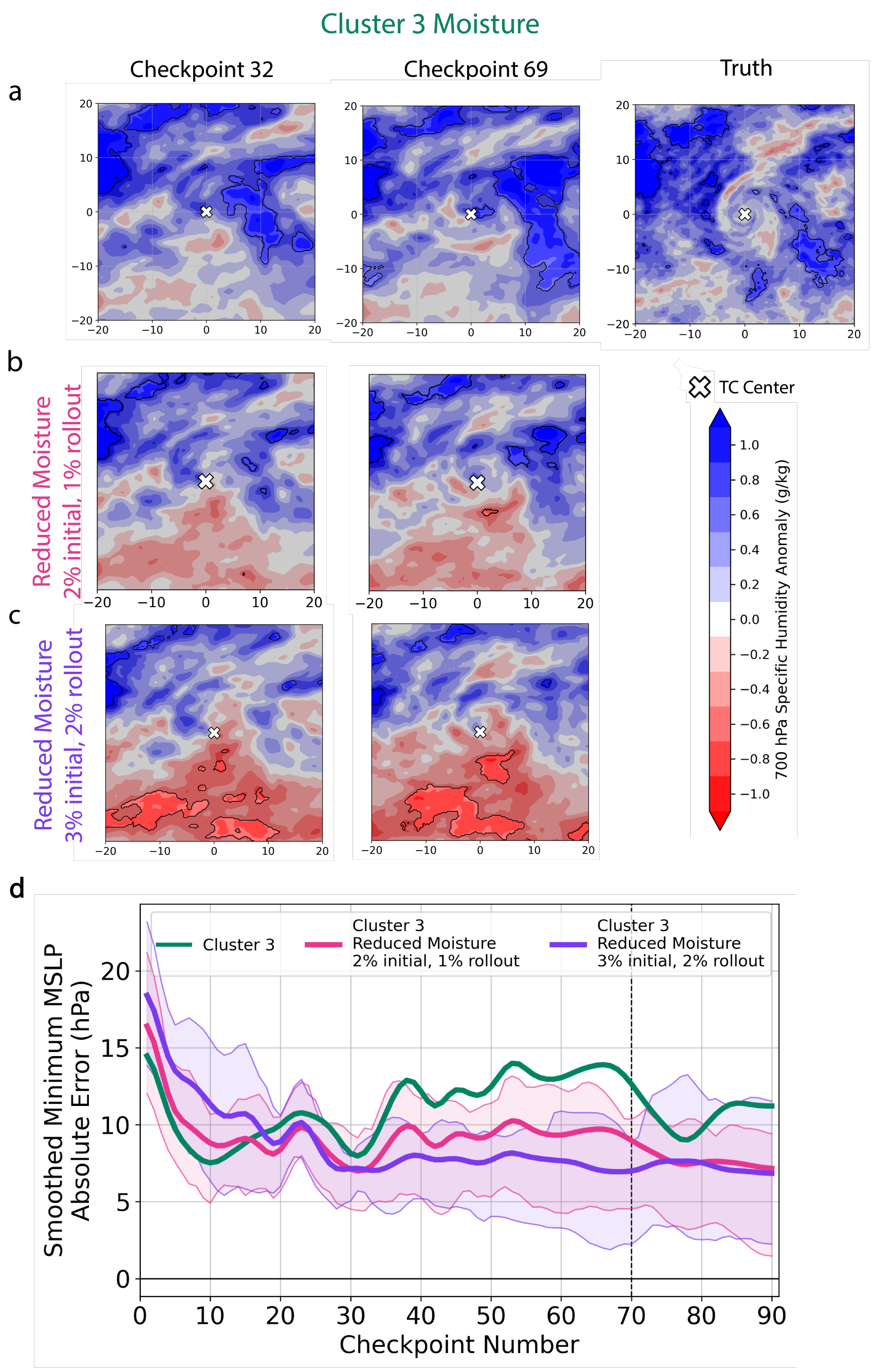

**Figure 3** Cluster 3 moisture experiment. Valid time Cluster 3 700 hPa specific humidity anomaly composites centered on the true TC center (white X) for Checkpoint 32, Checkpoint 69, and ERA5 truth (blue and red shading) **(a)**. Checkpoint 32 and checkpoint 69 forecasts for experiments with 2% initial/1% rollout **(b)** and 3% initial/2% rollout **(c)** moisture reductions. Anomalies are relative to the all-storm non-experimental forecast average for each checkpoint. Black outlines indicate significantly moist or dry subsets of all storms. Mean Cluster 3 minimum MSLP absolute error (dark green line, identical to Figure 2b) and reduced moisture experiments mean and interquartile range smoothed (Gaussian filter, sigma = 2) along checkpoints (pink and purple lines, shading) **(d).**

## 4 Discussion

We introduce the use of task-specific training dynamics for “watching” an AI weather emulator learn extreme weather through training. We find that the learning trajectories of SFNO five-day TC intensity forecasts are not homogeneous, and identify three distinct training patterns associated with these forecasts. A subset of about half of the TCs follows an expected learning trajectory with early checkpoints underestimating storm intensity, error decreasing through training and stabilizing towards the end resulting in skillful forecasts. Another subset of storms consistently underpredicts TC intensity, but shows minor improvement through training. The third subset of storms shows TC intensity improving at the beginning of training, but after checkpoint 30 there is a degradation in skill due to the SFNO overpredicting storm intensity. At first glance, loss of skill in the second half of training is consistent with SFNO blurring to avoid the double penalty problem, in which MSE-trained models smooth toward the conditional mean and suppress extreme values (Bonavita, 2024; Olivetti & Messori, 2024b). However, blurring would drive a weakening of all TCs through training and none of the three clusters show this trend: all three clusters show significant strengthening of TC intensity in the second half of training (Figure 2c).

These cases are not distinguished by storm characteristics (location, intensity, track, etc.) but they feature anomalously moist background environments. We conduct moisture reduction experiments that confirm a causal relationship between anomalous moisture and TC intensity skill degradation. However, the SFNO correctly forecasts this anomalous moisture, so the overprediction of storm intensity is not related to incorrect moisture forecasting. Instead, skill degradation must be a consequence of how the SFNO processes these samples. The model may over-rely on moisture as an indicator of TC intensification, or the anomalous moisture may place these cases in a region of the model’s learned distribution that systematically overpredicts TCs. Why anomalous moisture drives overprediction in these cases remains an open question with direct implications for the trustworthiness of AI weather emulators in extreme weather applications.

This study is one example of using task-specific training dynamics to build our understanding of how AI weather emulators learn. Assuming training checkpoints are saved (an inexpensive storage decision made at training time), this method does not require re-training models, making it relatively low cost. We see great potential in leveraging training dynamics for AI weather emulators, including experiments similar to our moisture experiments where one variable is perturbed (e.g. moisture) and task-specific training dynamics are evaluated (e.g. TC intensity) to identify when in training a model learns a physical relationship. In the exploding field of AI weather emulators, task-specific training dynamics allow us to connect AI learning and physical meteorology, which can improve AI models and, perhaps more importantly, gauge trust in existing models.

## Acknowledgments

This research was supported by Quadrature Climate Foundation Grant 01-21-000338 and Founder's Pledge award 240905. This research was supported by the Director, Office of Science, Office of Biological and Environmental Research of the US Department of Energy under contract no. DE-AC02-05CH11231 and by the Regional and Global Model Analysis Program area within the Earth and Environmental Systems Modeling Program. The research used resources of the National Energy Research Scientific Computing Center (NERSC), also supported by the Office of Science of the US Department of Energy, under contract no. DE-AC02-05CH11231. The computation for this paper was supported in part by the DOE Advanced Scientific Computing Research (ASCR) Leadership Computing Challenge (ALCC) 2023–2024 award "Huge Ensembles of Weather Extremes using the Fourier Forecasting Neural Network" to William Collins (LBNL) and ALCC 2024–2025 award "Huge Ensembles of Weather Extremes using the Fourier Forecasting Neural Network" to William Collins (LBNL).

## Availability Statement

Model weights for best SFNO checkpoint are stored at https://doi.org/10.5281/zenodo.17143349 (Baiman et al., 2025). Code and information for SFNO inference runs is available on DataDryad https://doi.org/10.5061/dryad.2rbnzs80n (Mahesh et al., 2025a). More information on training an SFNO model can be found at https://www.nvidia.com/en-us/high-performance-computing/earth-2/.

## Conflict of Interest Disclosure

The authors declare there are no conflicts of interest for this manuscript.

## References

Baiman, R., Barnes, E. A., & Mahesh, A. (2025). How Does an AI Weather Model Learn to Forecast Extreme Weather? [Data set]. Zenodo. https://doi.org/10.5281/ZENODO.17143349

Betley, J., Warncke, N., Sztyber-Betley, A., Tan, D., Bao, X., Soto, M., et al. (2026). Training large language models on narrow tasks can lead to broad misalignment. *Nature*, *649*(8097), 584–589. https://doi.org/10.1038/s41586-025-09937-5

Bi, K., Xie, L., Zhang, H., Chen, X., Gu, X., & Tian, Q. (2023). Accurate medium-range global weather forecasting with 3D neural networks. *Nature*, *619*(7970), 533–538. https://doi.org/10.1038/s41586-023-06185-3

Bonavita, M. (2024). On Some Limitations of Current Machine Learning Weather Prediction Models. *Geophysical Research Letters*, *51*(12), e2023GL107377. https://doi.org/10.1029/2023GL107377

Bonev, B., Kurth, T., Hundt, C., Pathak, J., Baust, M., Kashinath, K., & Anandkumar, A. (2023, June 6). Spherical Fourier Neural Operators: Learning Stable Dynamics on the Sphere. arXiv. https://doi.org/10.48550/arXiv.2306.03838

Charlton-Perez, A. J., Dacre, H. F., Driscoll, S., Gray, S. L., Harvey, B., Harvey, N. J., et al. (2024). Do AI models produce better weather forecasts than physics-based models? A quantitative evaluation case study of Storm Ciarán. *Npj Climate and Atmospheric Science*, *7*(1), 93. https://doi.org/10.1038/s41612-024-00638-w

DeMaria, M., Franklin, J. L., Chirokova, G., Radford, J., DeMaria, R., Musgrave, K. D., & Ebert-Uphoff, I. (2025). An Operations-Based Evaluation of Tropical Cyclone Track and Intensity Forecasts from Artificial Intelligence Weather Prediction Models. *Artificial Intelligence for the Earth Systems*, *4*(4), 240085. https://doi.org/10.1175/AIES-D-24-0085.1

Eyring, V., Collins, W. D., Gentine, P., Barnes, E. A., Barreiro, M., Beucler, T., et al. (2024). Pushing the frontiers in climate modelling and analysis with machine learning. *Nature Climate Change*, *14*(9), 916–928. https://doi.org/10.1038/s41558-024-02095-y

Hersbach, H., Bell, B., Berrisford, P., Hirahara, S., Horányi, A., Muñoz-Sabater, J., et al. (2020). The ERA5 global reanalysis. *Quarterly Journal of the Royal Meteorological Society*, *146*(730), 1999–2049. https://doi.org/10.1002/qj.3803

Hu, M. Y., Jain, S., Chaulagain, S., & Saphra, N. (2025, April 28). How to visualize training dynamics in neural networks. Retrieved from https://d2jud02ci9yv69.cloudfront.net/2025-04-28-visualizing-training-87/blog/visualizing-training/

Intergovernmental Panel On Climate Change (IPCC). (2023). *Climate Change 2021 – The Physical Science Basis: Working Group I Contribution to the Sixth Assessment Report of the Intergovernmental Panel on Climate Change* (1st ed.). Cambridge University Press. https://doi.org/10.1017/9781009157896

Kangaslahti, S., Rosenfeld, E., & Saphra, N. (2026, February 27). Hidden Breakthroughs in Language Model Training. arXiv. https://doi.org/10.48550/arXiv.2506.15872

Knapp, K. R., Kruk, M. C., Levinson, D. H., Diamond, H. J., & Neumann, C. J. (2010). The International Best Track Archive for Climate Stewardship (IBTrACS): Unifying Tropical Cyclone Data. *Bulletin of the American Meteorological Society*, *91*(3), 363–376. https://doi.org/10.1175/2009BAMS2755.1

Lam, R., Sanchez-Gonzalez, A., Willson, M., Wirnsberger, P., Fortunato, M., Alet, F., et al. (2023). Learning skillful medium-range global weather forecasting. *Science*, *382*(6677), 1416–1421. https://doi.org/10.1126/science.adi2336

Mahesh, A., Collins, W., Bonev, B., Brenowitz, N., Cohen, Y., Harrington, P., et al. (2025a). Huge ensembles part I design of ensemble weather forecasts with spherical Fourier neural operators; Huge ensembles part II properties of a huge ensemble of hindcasts generated with spherical Fourier neural operators (Version 11) [Data set]. Dryad. https://doi.org/10.5061/DRYAD.2RBNZS80N

Mahesh, A., Collins, W., Bonev, B., Brenowitz, N., Cohen, Y., Elms, J., et al. (2025b, April 3). Huge Ensembles Part I: Design of Ensemble Weather Forecasts using Spherical Fourier Neural Operators. https://doi.org/10.48550/arXiv.2408.03100

Mahesh, A., Collins, W., Bonev, B., Brenowitz, N., Cohen, Y., Harrington, P., et al. (2025c, April 3). Huge Ensembles Part II: Properties of a Huge Ensemble of Hindcasts Generated with Spherical Fourier Neural Operators. https://doi.org/10.48550/arXiv.2408.01581

NVIDIA. (n.d.). Makani. Retrieved from https://github.com/NVIDIA/makani

Olivetti, L., & Messori, G. (2024a). Advances and prospects of deep learning for medium-range extreme weather forecasting. *Geoscientific Model Development*, *17*(6), 2347–2358. https://doi.org/10.5194/gmd-17-2347-2024

Olivetti, L., & Messori, G. (2024b). Do data-driven models beat numerical models in forecasting weather extremes? A comparison of IFS HRES, Pangu-Weather, and GraphCast. *Geoscientific Model Development*, *17*(21), 7915–7962. https://doi.org/10.5194/gmd-17-7915-2024

Pasche, O. C., Wider, J., Zhang, Z., Zscheischler, J., & Engelke, S. (2025). Validating Deep Learning Weather Forecast Models on Recent High-Impact Extreme Events. *Artificial Intelligence for the Earth Systems*, *4*(1), e240033. https://doi.org/10.1175/AIES-D-24-0033.1

Sun, Y. Q., Hassanzadeh, P., Zand, M., Chattopadhyay, A., Weare, J., & Abbot, D. S. (2025). Can AI weather models predict out-of-distribution gray swan tropical cyclones? *Proceedings of the National Academy of Sciences*, *122*(21), e2420914122. https://doi.org/10.1073/pnas.2420914122

Tirumala, K., Markosyan, A. H., Zettlemoyer, L., & Aghajanyan, A. (2022, November 2). Memorization Without Overfitting: Analyzing the Training Dynamics of Large Language Models. arXiv. https://doi.org/10.48550/arXiv.2205.10770

Watt-Meyer, O., Henn, B., McGibbon, J., Clark, S. K., Kwa, A., Perkins, W. A., et al. (2025). ACE2: accurately learning subseasonal to decadal atmospheric variability and forced responses. *Npj Climate and Atmospheric Science*, *8*(1), 205. https://doi.org/10.1038/s41612-025-01090-0

Supporting Information for

# Watch an AI Weather Model Learn (and Overshoot) Tropical Cyclones

**Rebecca Baiman[1,2], Ankur Mahesh[3,4], and Elizabeth A. Barnes[1,2,5]**

[1]Department of Atmospheric Science, Colorado State University, Fort Collins, CO, USA.

[2]Faculty of Computing and Data Science, Boston University, Boston, MA, USA

[3]Earth and Environmental Sciences Area, Lawrence Berkeley National Laboratory (LBNL), Berkeley, CA, USA.

[4]Department of Earth and Planetary Science, University of California, Berkeley, CA, USA

[5]Department of Earth and Environment, Boston University, Boston, MA, USA

**Contents of this file**

**Introduction**

Text S1 describes metrics in assessing K-means clusters in main article. Supplementary figures referenced in the main article.

## Text S1.

We use k-means clustering to group 106 TC intensity forecast – truth training curves into clusters. After standardizing the data, we use the elbow method and silhouette scores to identify k=3 clusters (Figure S1a and b). Based on both elbow method and silhouette scores, we find that 2 or 3 clusters are reasonable for this dataset. Even though k=2 produces a higher silhouette score, we choose the second-best k=3 to identify an additional training dynamic behavior. We conduct a sensitivity analysis with k=2 and find one cluster that underpredicts TC intensity and one that overpredicts on average. The overpredicting cluster has a significantly worse forecast at checkpoints 65–70 compared to checkpoints 28–33, consistent with the manuscript Cluster 3 but weaker in magnitude. The overpredicting cluster does not feature significantly more moisture, implying that the Cluster 3 used in this study represents a subset of overpredicted TCs that are associated with moist environments. We are confident based on our moisture experiment results that Cluster 3 represents a meaningful grouping of overpredicted storms.

The k-means clustering into 3 groups can be visualized using the first two principal components (Figure S1c). We note that the clustering is well represented in the first principal component, which explains 69% of the variance between TC training curves. Inspecting the weights of this first principal component, we find that the highest weights occur near the end of training (not shown). Thus, the first principal component is capturing the storm training dynamic variance towards the end of training.

# K-Means Methodology

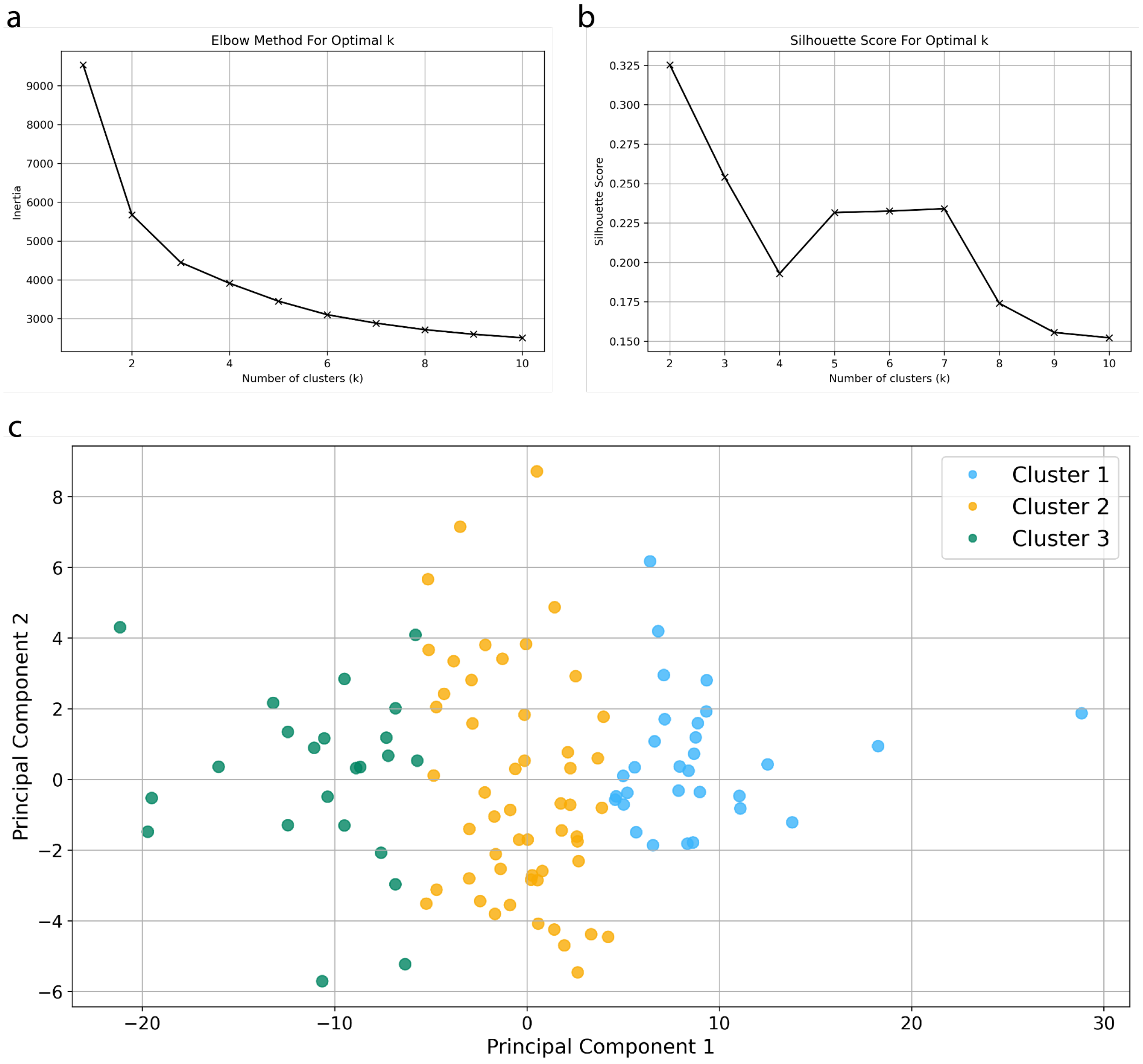

**Figure S1.** Inertia (sum of the squared distances between each learning trajectory and the center of the cluster) (**a**) and silhouette score (a measure of the distance between points in the same cluster versus points in the closest other cluster, closer to 1 is better) (**b**) for k-means

clusters with k =2–10. Clusters 1–3 TC learning tractories of minimum MSLP forecast – truth plotted in two dimensions using principal component analysis **(c).**

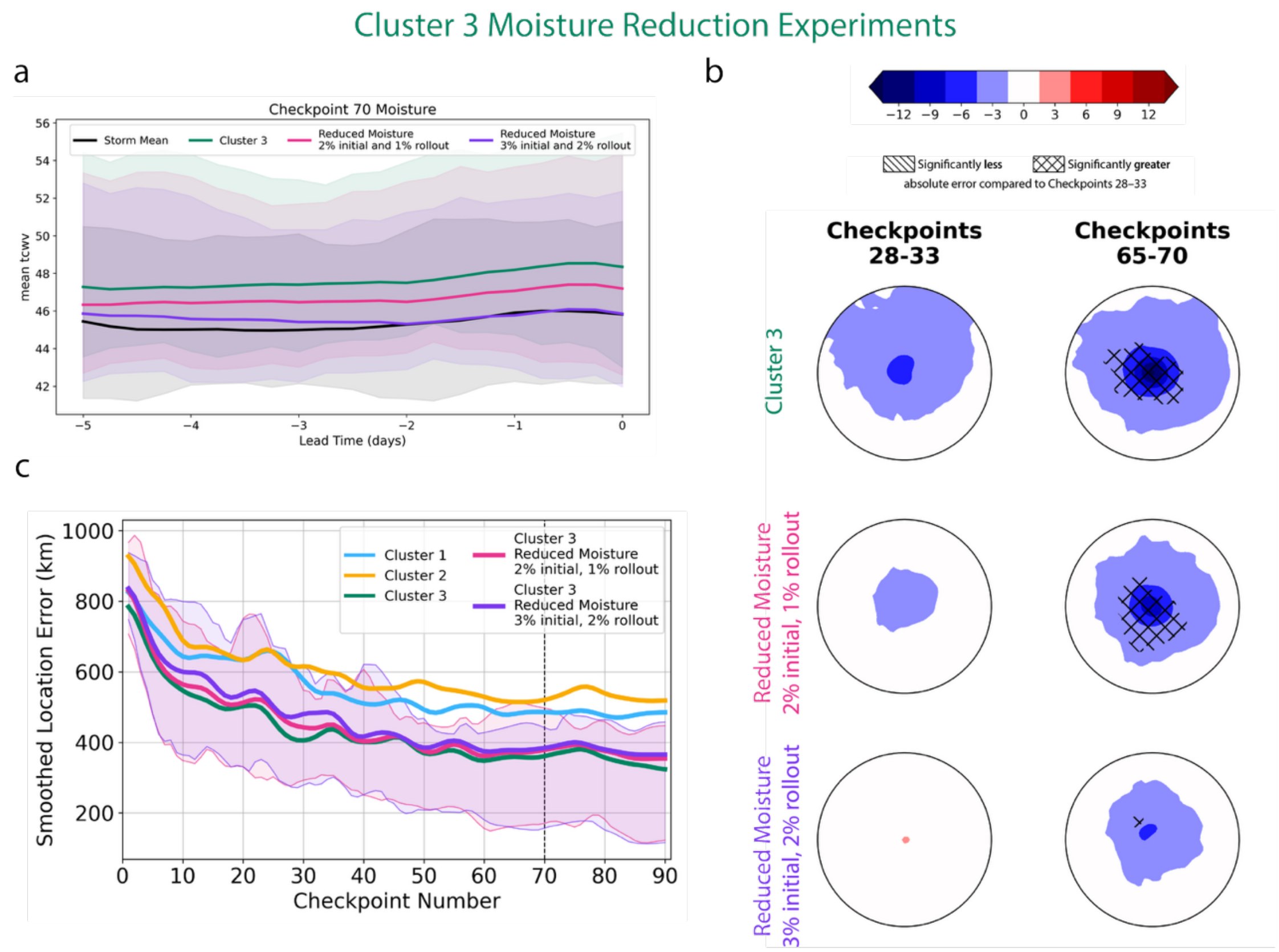

**Figure S2** Checkpoint 70 mean storm domain total column water vapor mean and interquartile range through the five-day forecast for Cluster 3 storms (green), reduced moisture experiments (pink and purple) and all storms (black) **(a).** Minimum MSLP centered composites for storms in Clusters 1–3 (rows) and checkpoints 28–33 (left column) and checkpoints 65–70 (right column) (shading, **b**). Significantly greater (smaller) checkpoint 67–70 values are determined using a t test for difference in means with a p-value<.05 and where the storm-centered MSLP Forecast minus Truth difference is at least 3 hPa (stripes and hatching, **b**). Mean for each cluster of absolute value of forecast minus truth location error smoothed along checkpoints for each storm (Gaussian filter, sigma = 2) **(blue, yellow, green lines c).** Mean and interquartile range of reduced moisture experiments absolute value of forecast minus truth minimum MSLP smoothed along checkpoints for each storm (Gaussian filter, sigma = 2) **(pink and purple c).**

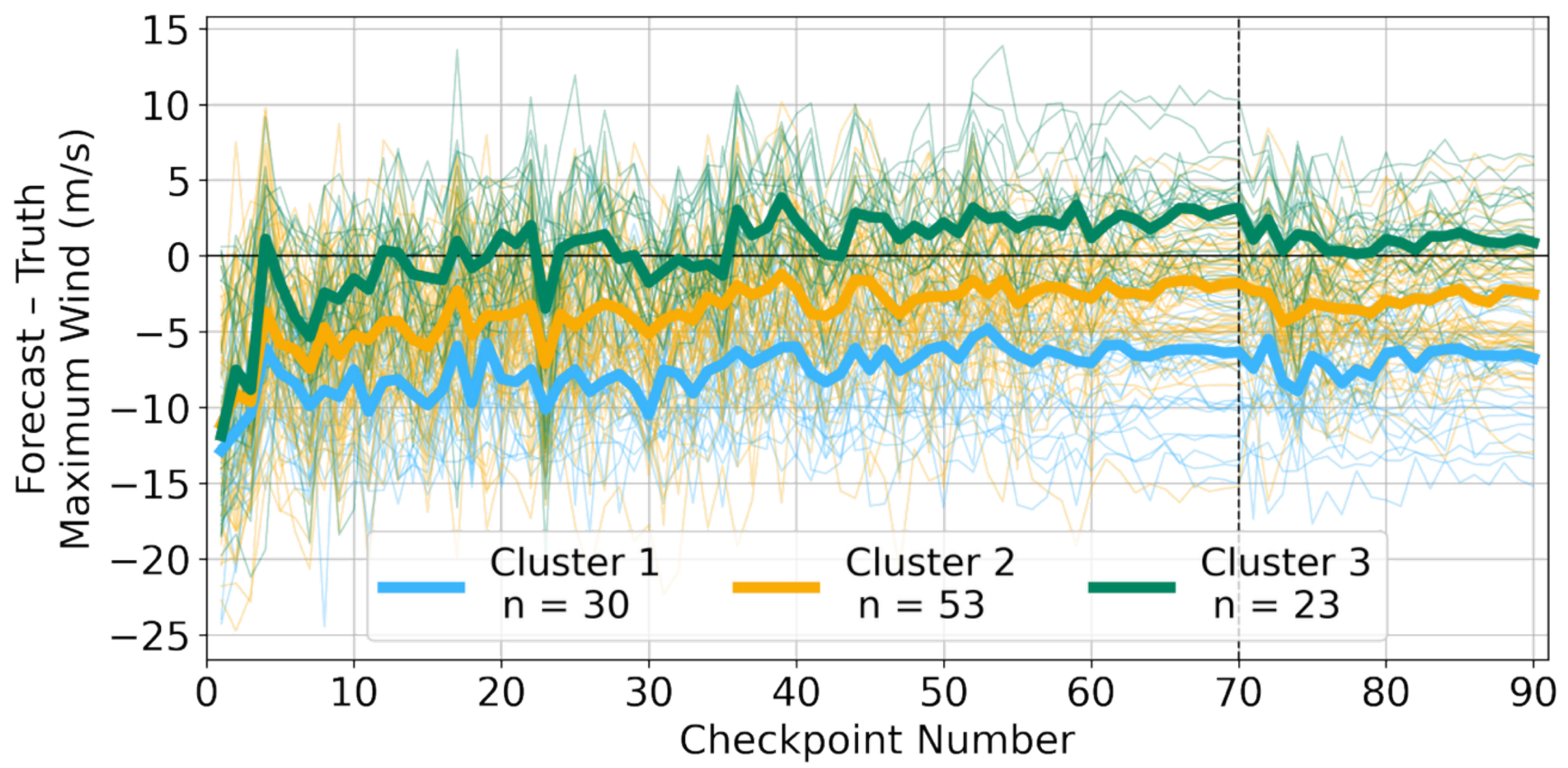

**Figure S3** Training dynamics of 106 TCs**.** TC forecast minus truth maximum 10m wind speed for each storm (thin lines) and mean for each cluster (thick lines) colored by cluster.

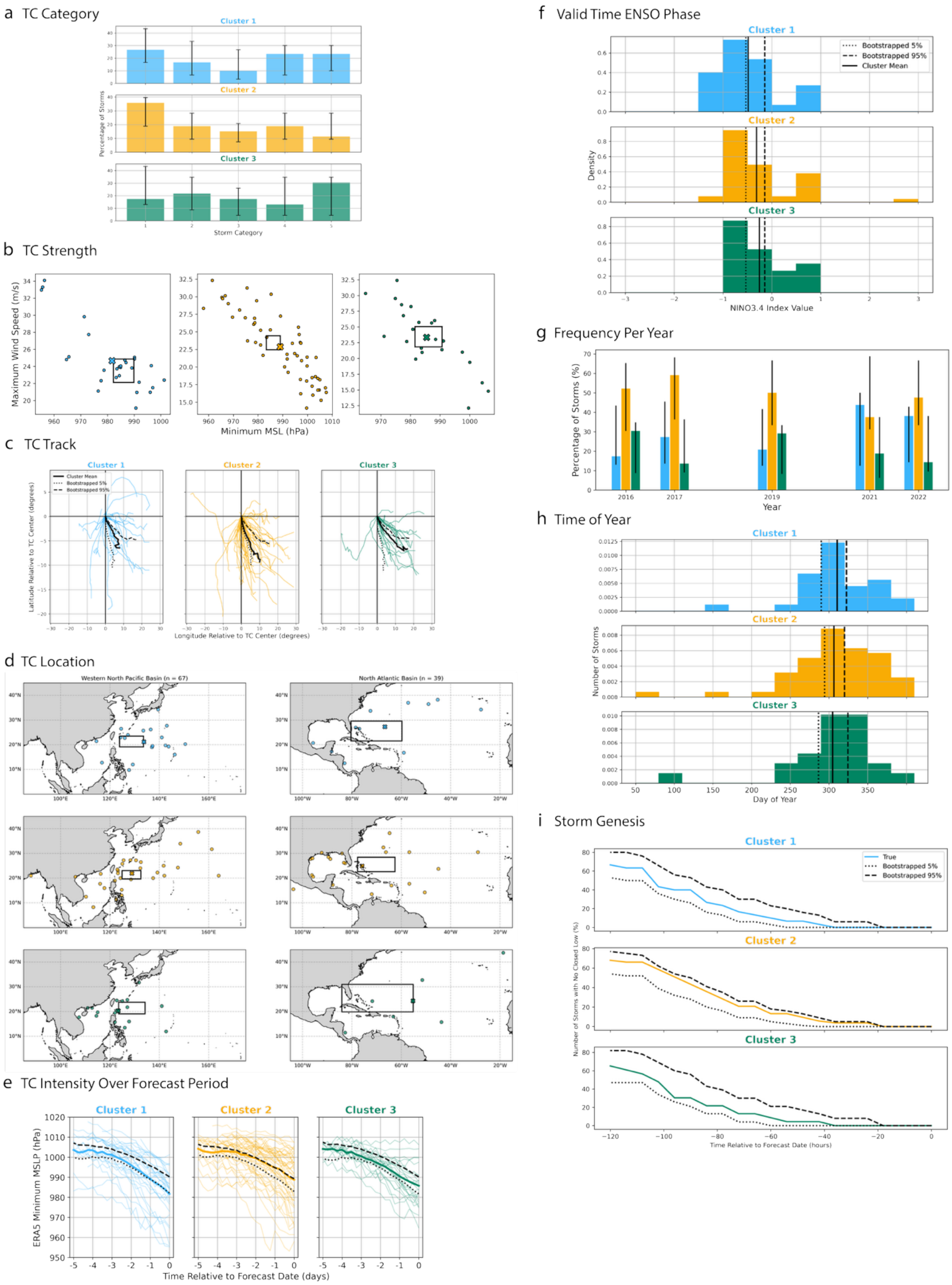

**Figure S4.** TC characteristics by cluster. Significance is assessed using a bootstrapping method. We compare each cluster value to a bootstrapped distribution of 1,000 random samples (with size n=number of samples in cluster) drawn from the collection of 106. A cluster is considered a significantly unique subset of storms if the cluster value falls outside of the 5–95% spread of the bootstrapped distribution. This spread is represented by dotted and dashed lines, respectively, unless otherwise noted.

Percentage of storms in each TC category for each cluster. Black vertical lines show 5-95% range of bootstrapped samples **(a)**. ERA5 minimum MSLP and maximum wind speed of each storm (circular points) and mean (x) for each cluster. Black boxes show the 5-95% range of bootstrapped samples **(b)**. ERA5 storm track from initialized timestep to valid time 5 days later for each storm (thin lines) and cluster mean (solid black line). Tracks are centered on the valid time location of the storm, and track is only calculated for timesteps that feature a closed MSLP low **(c)**. TC valid time location in ERA5, colored by cluster. Black boxes show the latitude and longitude 5-95% range of bootstrapped samples **(d)**. ERA5 minimum MSLP over the forecast period for each storm (thin lines) and mean (bold line) for each cluster **(e)**. Distribution of storms in each cluster by the Niño3.4 Index. Solid black line shows the mean for each cluster (**f**). Frequency of each cluster by year. Black vertical lines show 5-95% range of bootstrapped samples (**g**). Distribution of storms in each cluster by day of year. The mean is shown with a solid black line (**h**). Number of storms with no closed low in the ERA5 MSLP field over forecast period (**i**).

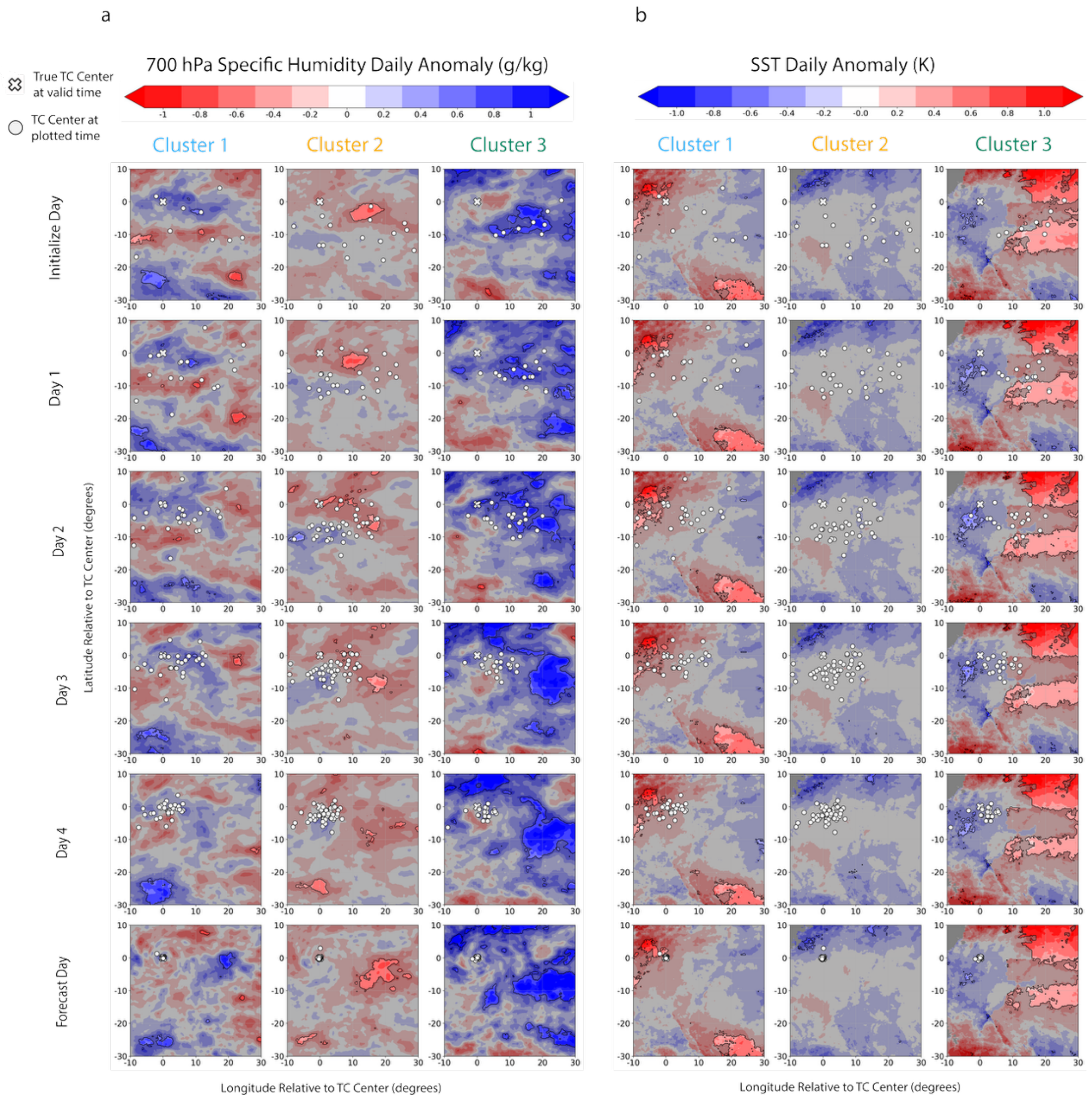

**Figure S5.** ERA5 daily anomalies of 700 hPa specific humidity (**a**) and SST (**b**) starting from the day of initialization (top row) to the forecasted day (bottom row). Centering is based on true location of minimum MSLP at the valid time (white x). Cluster anomalies are compared to all storms. Areas where cluster anomalies are a significant subset of all storms are outlined in black. Significance is calculated using a bootstrapping method where significant values must fall in the top or bottom 5th percentile compared to a distribution of bootstrapped samples (size n to match each cluster) from all storms.

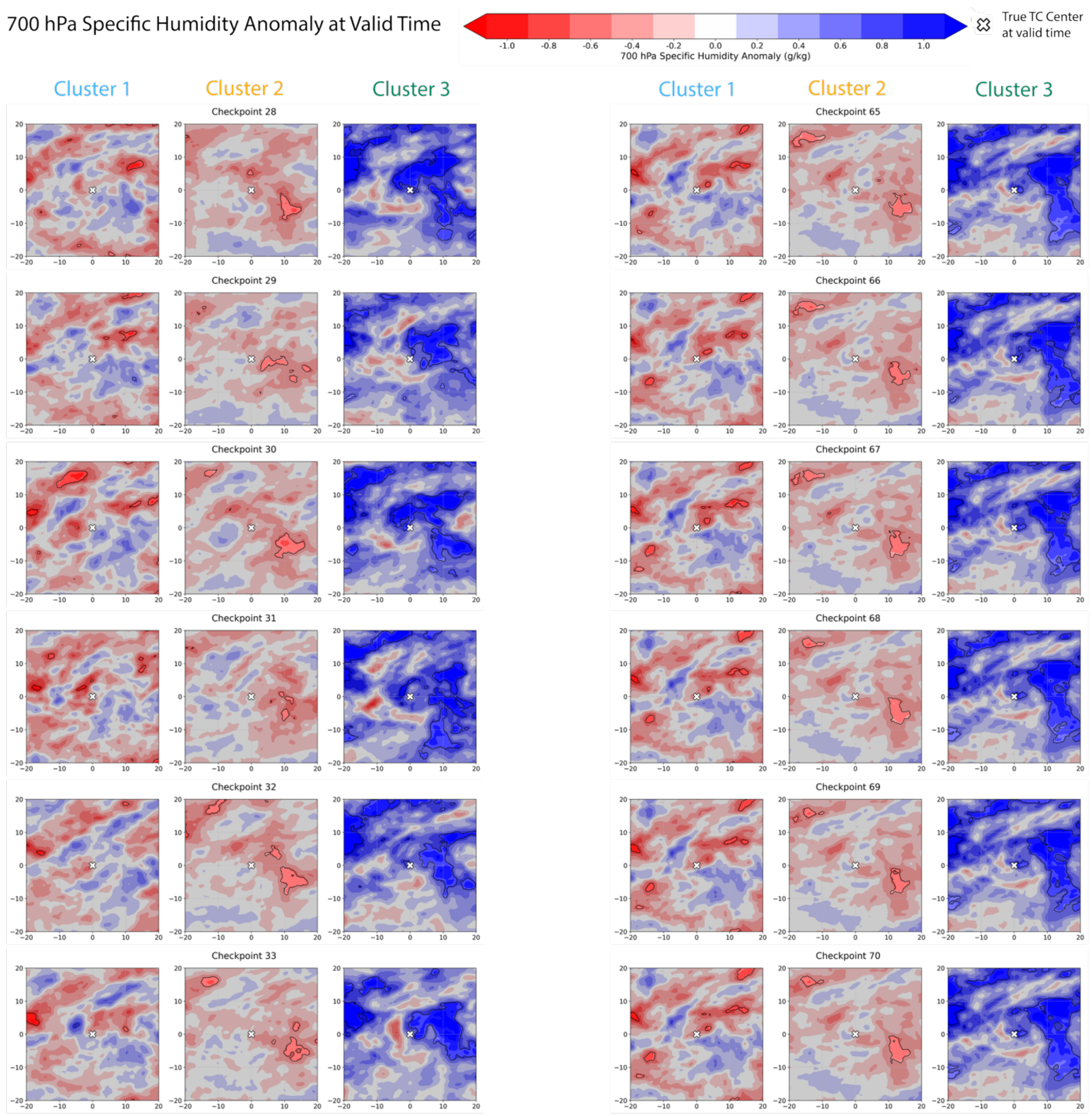

**Figure S6.** SFNO forecast anomalies of 700 hPa specific humidity for checkpoints 28–33 and 65–70. Centering is based on true location of minimum MSLP at the valid time (white x). Cluster anomalies are compared to all storms. Areas where cluster anomalies are a significant subset of all storms are outlined in black. Significance is calculated using a bootstrapping method where significant values must fall in the top or bottom 5th percentile compared to a distribution of bootstrapped samples (size n to match each cluster) from all storms.